\documentclass[3p,times,authoryear]{elsarticle}

\usepackage{amsmath,amsfonts,amssymb,amsthm}

\usepackage{mathtools}
\usepackage{multirow}
\usepackage{unitsdef}
\usepackage{xcolor}
\usepackage[left]{lineno} 
\usepackage{hyperref}
\usepackage{diagbox}
\modulolinenumbers[1]

\DeclareMathOperator*{\argmin}{arg\,min}

\usepackage{setspace}
\journal{Journal of Geophysical Research: Solid Earth}

 \biboptions{authoryear}

\definecolor{myc}{RGB}{0, 63, 152}
\begin{document} 
	
	\begin{frontmatter}
		\title{Implicit full waveform inversion with deep neural representation}
		
		\author[addressA,addressB]{Jian Sun\corref{CA}}  
		\author[addressC]{Kristopher Innanen}  

		\address[addressA]{College of Marine Geosciences, Ocean University of China}
		\address[addressB]{Key Lab of Submarine Geosciences and Prospecting Techniques MOE China, Ocean University of China}
		\address[addressC]{CREWES Project, Department of Geosciences, University of Calgary}
		\cortext[CA]{Corresponding author}
		\ead{jiansun@ouc.edu.cn}
		
		\begin{abstract}
			Full waveform inversion (FWI) commonly stands for the state-of-the-art approach for imaging subsurface structures and physical parameters, however, its implementation usually faces great challenges, such as building a good initial model to escape from local minima, and evaluating the uncertainty of inversion results. In this paper, we propose the implicit full waveform inversion (IFWI) algorithm using continuously and implicitly defined deep neural representations. Compared to FWI, which is sensitive to the initial model, IFWI benefits from the increased degrees of freedom with deep learning optimization, thus allowing to start from a random initialization, which greatly reduces the risk of non-uniqueness and being trapped in local minima. Both theoretical and experimental analyses indicates that, given a random initial model, IFWI is able to converge to the global minimum and produce a high-resolution image of subsurface with fine structures. In addition, uncertainty analysis of IFWI can be easily performed by approximating Bayesian inference with various deep learning approaches, which is analyzed in this paper by adding dropout neurons. Furthermore, IFWI has a certain degree of robustness and strong generalization ability that are exemplified in the experiments of various 2D geological models. With proper setup, IFWI can also be well suited for multi-scale joint geophysical inversion.
		\end{abstract}
			
	\end{frontmatter}

	
\section{Introduction}
Imaging high-resolution heterogeneities of the Earth subsurface is the key objective of geophysical inversion, which is usually performed in the manner of an iterative optimization procedure by minimizing the misfit between simulated data and observations. With appropriate data acquisitions and approximations, full waveform inversion (FWI) commonly stands for the state of the art technique in geophysical inversion due to its maximized utilization of the wavefield information \citep{tarantola1984inversion}. Under certain physical assumptions, FWI repetitively reduces the discrepancy measured between sampled solutions of wave propagation problems, which are obtained by solving the wave equation within an initial or intermediate subsurface model, and local measurements of geophysical wavefield \citep{virieux2009overview}. 

The optimization procedure of FWI is based on the assumption of convexity, which is inconsistent with the non-linearity of its objective function, and normally carried out with an adjoint state method \citep[][]{plessix2006review,liu2006finite,yedlin2010tutorial}. The ill-posedness and non-linearity of FWI pose great challenges to its optimization process, which are mainly manifested in the form of cycle-skipping. Generally, this can be reduced by either involving low-frequency/high-angle measurements of geophysical data or starting from a good initial model \citep{virieux2009overview}. However, the acquisition of low-frequency geophysical data remains to be a great challenge because of its susceptibility to noise. Thus, a common practice is formulating new objectives, such as regularization \citep{lin2014acoustic,esser2018total}, envelope-based function \citep{bozdaug2011misfit,chi2014full}, reflection waveform inversion \citep[e.g., RWI,][]{xu2012full,chi2015correlation}, and adaptive waveform inversion \citep[e.g., AWI,][]{warner2016adaptive,guasch2019adaptive}, to make FWI optimization focus on the high-angle characteristics as the substitution of low-frequency data. The other alternative solution is by building a realistic model using other methods as the starting point of FWI. For instance, to invert a smooth model, one can use traveltime tomography by matching the first-break information \citep{clement2001migration} or adopt migration velocity analysis by measuring the flatness of the common-image gathers \citep{symes2008migration}. In addition to the cycle-skipping issue, FWI in general is computationally demanding, which makes huge impacts on the implementation of current FWI algorithm \citep{byrd1995l-BFGS,nocedal2006cg,metivier2013full}. Besides that, evaluating the uncertainty of FWI is also an obstacle because of the high-dimensionality of the model space when considering large numbers of subsurface parameters.

Deep learning, primarily in the form of deep neural networks (DNNs), is having widespread attention in scientific and engineer fields, such as image classification/recognition, shape representation, self-driving, machine translation, and natural language processing, due to its increased ability to sense data and immense improvements of large-scale computational power. For instance, in shape representation, coordinate-based DNNs are trained to learn continuous signed distance functions from a single or multi-view images \citep{chen2019learning,mescheder2019occupancy,park2019deepsdf,mildenhall2020nerf,sitzmann2020implicit}. After training, such a DNN can be treated as an implicit representation of an image or a 3D object with respect to coordinates, which provides a new perspective for image/object reconstruction.

Tremendous efforts have been made to facilitate deep learning applications in geophysics and its practitioners, such as automatic first-arrival picking \citep{wu2019semiautomatic, cano2021automatic,yuan2020robust}, seismic facies classification \citep{alaudah2019machine,liu2020seismic,feng2021bayesian}, noise attenuation \citep{saad2020deep,yang2021deep,wu2019white} - domains for which data are plentiful and physical-law does not explicitly reveal. Deep learning has had surprising successes in such domains, in which problems to be solved can be treated/converted into the cognitive tasks of pattern recognition, by maximizing the utilization of data in the training aspect. Attempts have also been made to devise a DNN architecture that directly maps seismograms to subsurface models in a fully data-driven sense \citep{wang2018velocity,wu2018inversionnet,yang2019deep}. However, due to the ill-posedness and complexity of seismic inversion, training a well-generalized and robust DNN for inversion purpose requires vastly various pairs of subsurface models and their corresponding seismograms. Building such a dataset that contains all possible combinations of subsurface structures is nearly impossible to accomplish.

In comparison with the purely data-driven approaches, which are unlikely predictive, FWI has shown persistent predictability and formidable generalization ability with the guidance of governing equations. Thus, we must not omit the known underlying physical principles when pursing a new generation of data-centric deep learning inversion. With this in mind, several ways of incorporating FWI with DNNs have been explored. \cite{sun2020theory} propose a theory-guided seismic inversion framework in which the forward modeling is devised in a framework of recurrent neural network (RNN) and the inversion process is described as the RNN training using automatic differentiation that can be easily accelerated by deep learning platforms. The theory-guided RNN framework can also be extended to more complete multidimensional parameter estimations \citep{zhang2020numerical} or electromagnetic inversion \citep{hu2021theory}. \cite{wu2019parametric} and \cite{he2021reparameterized} parameterize velocity model with a convolutional neural network (CNN) that creates a multi-grid representation of velocity and automatically added regularization effects in the FWI procedure. \cite{sun2021physics} develop a hybrid network design by simultaneously involving a data-driven model misfit and a physics-guided data residual during the training of the network. While these work are creative, they still encounter major challenges as traditional FWI does. For instance, they either require good initial models to pretrain the CNN or have difficulty to evaluate the uncertainty of inversion results.

The goal of this paper is to further explore deep fusion of FWI and DNNs in a way that offers the possibility to address the challenges of FWI in initial model building and uncertainty analysis, while retaining the strong robustness and generalization ability of FWI. According to the universal approximation theorem, DNNs are able to represent any continuous functions. For instance, in shape representation, the surface of the target object can be interpreted as a continuous signed distance function. Analogously, we are able to build an implicit functional space using DNNs that represent various parameterizations of the subsurface model. Comparing to the CNN reparameterization, which lacks the flexibility (i.e., supports only fixed-size outputs and has the difficulty in transferring to other models) and requires initial models for pretraining, the implicit function built with DNNs can be adapt to physical parameterizations and models of arbitrary size, due to the properties of continuous functions. In addition, we also seek an implicit function using DNNs to reduce the dependency of FWI on the initial model by increasing the numbers of degree of freedoms in seismic inversion.

This paper is organized as follows. First, we introduce the concept of deep neural representation (DNR), and then demonstrate how to build FWI with implicit neural representation. Second, we discuss the network selections for DNR and examine its capacity of representing subsurface geological models. After that, experimental examples are presented to exemplify the performance of the proposed method, including random initialization, robustness, uncertainty evaluation, and generalization ability. Finally, we discuss its further potentials as well as challenges.

\section{Methodology}
\subsection{Deep Neural Representation}

Assume we are interested in a set of features $\Psi$ that can be interpreted as a continuous function $\Phi$, with respect to the input $\mathbf{x} \in \Omega_k, k=1, ..., K$, which is implicitly defined by equations of the form
\begin{equation} \label{eqn:genFunc}
	\mathcal{L}_k(\mathbf{x}, \Phi, \nabla_{\mathbf{x}} \Phi, \nabla_{\textbf{x}}^2 \Phi, ...) = 0, \quad \Phi: \mathbf{x} \rightarrow \Phi(\mathbf{x})
\end{equation}
where the derivation of $\mathcal{L}_k$ can be purely data-driven or physics-deterministic and the physical meaning of the input $\mathbf{x}$ relys on features of interest $\Psi$ to be represented.

 Based on the universal approximation theorem, there must exist a neural network that is equipped to map $\mathbf{x}$ to the quantity of interest $\Psi$ while satisfying constraints shown in equation~\ref{eqn:genFunc}. Learning a network that parameterizes an implicitly defined function $\Phi$ is referred to as DNR. The training process of such a network denoted as $\mathcal{N}_{\Theta}$ can be implemented by minimizing
\begin{equation} \label{eqn:costFunc}
	\argmin_{\Theta} \sum^K_{k=1} \lambda_k ||\mathcal{L}_k(\mathbf{x}, \mathcal{N}_{\Theta}, \nabla_{\mathbf{x}} \mathcal{N}_{\Theta}, \nabla_{\textbf{x}}^2 \mathcal{N}_{\Theta}, ...)||^2, \quad \mathcal{N}_{\Theta}: \mathbf{x} \rightarrow \mathcal{N}_{\Theta} (\mathbf{x})
\end{equation}
where $\Theta$ are the trainable weights and biases of the neural network, $\lambda_k$ denotes the trade-off parameter.

With proper training, the DNR-based network $\mathcal{N}_{\Theta}$ is approximately equivalent to the implicitly defined function $\Phi$ for the input domains of $\mathbf{x}$, i.e.,  $\mathcal{N}_{\Theta}(\mathbf{x}) \approx \Phi(\mathbf{x}), \forall \mathbf{x} \in \Omega_k$. It indicates that we may use a continuous and differentiable function, instead of a given vast and chaotic patterns, to represent features of interest for a certain input domain. Comparing to the discrete parameterization, in which memory and precision are highly dependent on the grid resolution, a continuous representation $\Phi(\mathbf{x})$/$\mathcal{N}_{\Theta}(\mathbf{x})$ on the continuous domain of $\mathbf{x}$ can be much more efficient while preserving fine details. For instance, instead of saving tremendous grid-based wavefields, a better alternative is saving the continuous implicit representation of wavefields constrained by wave equations, which can be commonly solved in the framework of physics-informed neural networks (PINNs). Thus, one can say that PINNs are special forms of DNR, in which constraints $\mathcal{L}_k$ are determined by the underlying physical principles and the input domains are usually defined as the spatial or spatio-temporal coordinates.

\subsection{Implicit Full Waveform Inversion}
The formalization of DNR can also be well suited to a wide variety of problems in the scientific and engineering fields. For instance, in geophysical inverse problems, the features of interest $\Psi$ are physical properties of the subsurface model denoted as $\mathbf{m}$, where the constraints $\mathcal{L}$ can be determined by wave propagation theory in FWI or by Zoeppritz equation in amplitude-versus-offset (AVO) inversion. we refer to FWI using implicit deep neural representation as implicit full waveform inversion, denoted as IFWI. In the case of IFWI, equation~\ref{eqn:genFunc} can be rewritten as
\begin{equation} \label{eqn:ifwiFunc}
	\mathbf{R} \mathcal{F}(\mathbf{m}, \mathbf{s}, \mathbf{x}, \mathbf{t}) - \mathbf{d} = 0, \quad \mathbf{m}: \mathbf{x} \rightarrow \mathbf{m}(\mathbf{x})
\end{equation}
where $\mathcal{F}$ denotes the forward modeling operator of wave propagation, $\mathbf{x}$ and $\mathbf{t}$ are spatial and temporal coordinates, respectively, $\mathbf{s}$ represents the source information, $\mathbf{R}$ denotes the matrix of receiver layouts, and $\mathbf{d}$ is the observed data.

The optimal implicit representation are obtained by minimizing the objective function of IFWI, using
\begin{equation} \label{eqn:ifwi-objFunc}
	\argmin_{\Theta} ||\mathbf{R} \mathcal{F}(\mathcal{N}_{\Theta}, \mathbf{s}, \mathbf{x}, \mathbf{t}) - \mathbf{d} ||^2,  \quad \mathcal{N}_{\Theta}: \mathbf{x} \rightarrow \mathcal{N}_{\Theta}(\mathbf{x}) \approx \mathbf{m(x)}
\end{equation}

Equation~\ref{eqn:ifwi-objFunc} indicates that we are seeking for a continuous and implicit functional representation, instead of a grid-based solution, of subsurface parameters using IFWI. Note that IFWI allows its optimization to be performed in a mesh-free manner as long as a mesh-free forward operator $\mathcal{F}$ (for instance, with PINNs solver) is adopted. However, to concentrate on the validity of DNR, a time-domain grid-based forward modeling is employed using finite difference method in this paper. Moreover, the forward simulation of wave propagation is embedded into the theory-designed RNN, as shown in the work of \citep{sun2020theory}, to make the best use of deep learning platform.

\subsection{Network Design}
Multilayer perceptrons (MLPs) have shown its great potentials in DNR fields, such as shape representation \citep{chen2019learning, genova2019learning,park2019deepsdf}, object reconstruction \citep{mescheder2019occupancy,xie2019pix2vox}, and scene representation \citep{mildenhall2020nerf,sitzmann2019scene}. Due to its continuous and memory-efficient characteristics, we select MLPs as the primary architecture for IFWI. The neural network to represent the subsurface model takes the spatial coordinates (or spatio-temporal coordinates in the time-lapse IFWI) as input and outputs the physical parameters.

To learn a continuous and implicit function of the complexly distributed subsurface parameters, MLPs must be equipped with the nonlinear activation functions. The commonly suited activation functions are sigmoid, hyperbolic tangent (tanh), rectified linear unit (ReLU) and its variants, of which ReLU is well known and widely used as the simplicity of its derivatives that prevents the gradient vanishing during the backpropagation of DNNs.
However, \cite{sitzmann2020implicit} demonstrate that MLPs with these non-periodic activation functions have difficulties in learning high-frequency components and high-order derivatives of images or scenes to be represented. Inspired by discrete cosine transform \citep{klocek2019hypernetwork}, \cite{sitzmann2020implicit} propose MLPs with periodic activation functions, specially MLPs with sinusoidal functions (also known as sinusoidal representation network, denoted as SIREN), and demonstrate that the periodic characteristic of activation functions offer surprising benefits in learning high-frequency and high-order derivatives information over these non-periodic activation functions.

To re-exemplify the merits of using periodic activation functions in subsurface representation, we build a SIREN to reconstruct the physical parameterizations of the 2D Marmousi model from the spatial coordinates in the horizontal and depth directions. To enlarge the output range of the proposed SIREN, which is limited by the sinusoidal activation function, a linear output layer is selected, i.e., no activation function is applied to the output layer. Furthermore, instead of directly exporting the compressional wave velocity $Vp$, the SIREN outputs its normalization. Here, the mean and standard deviation of compressional wave velocity of the Marmousi model are utilized to perform the normalization.

MLPs with four hidden layers, each of which contains 128 neurons, are selected for subsurface parameterizations of the acoustic Marmousi model.  Proposed MLPs are equipped with sinusoidal activation functions and ReLUs, respectively, where layers of the ReLU-based MLP are initialized using a uniform distribution $\omega_i \sim \mathcal{U}(-\sqrt{1/n}, \sqrt{1/n})$ and layers of the SIREN are also uniformly initialized but with an alternative bound $\omega_i \sim \mathcal{U}(-\sqrt{6/n}, \sqrt{6/n})$ by following the principled scheme in the work of \cite{sitzmann2020implicit}. Here, $n$ indicates the number of input features fed into the layer being initialized. Besides that, to further accelerate the frequency learning throughout the SIREN, a fixed tuning weight $\omega_0=30$ is added into the sinusoidal activation functions $\sin(\omega_0 \cdot \mathbf{W} \mathbf{x} + \mathbf{b})$ of the first hidden layer. Refer to \cite{sitzmann2020implicit} for more details on initialization of the SIREN. A variant of the gradient decent algorithm, adaptive momentum (Adam) optimizer, is selected to train both MLPs with whose respective optimal learning rates discovered through extensive trial and error runs (in this case, $1\times10^{-3}$ for the ReLU-based MLP and $1\times10^{-4}$ for the SIREN).

\begin{figure}[ht]
	\centering
	\includegraphics[width=\textwidth]{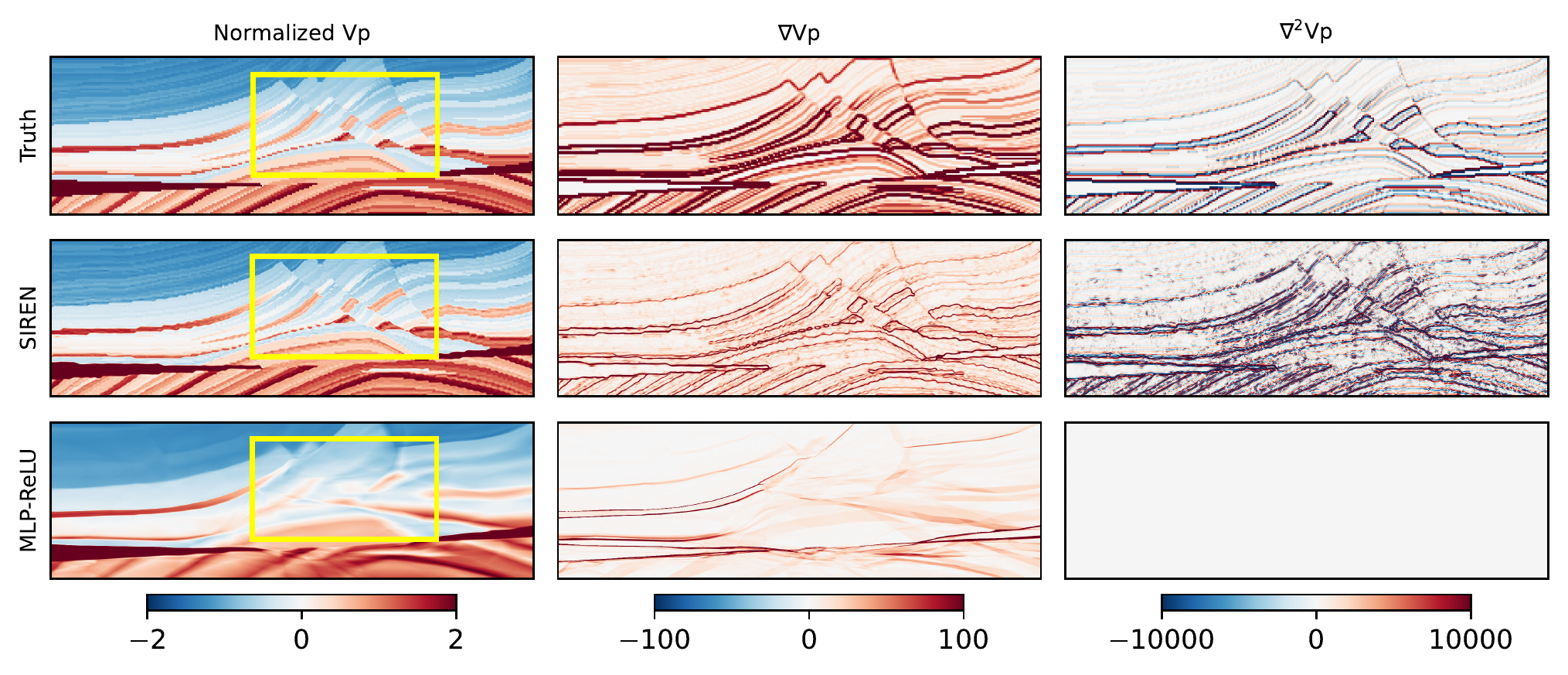}
	\caption{Comparison of implicit neural representation using different activation functions to fit a 2D acoustic Marmousi model. Columns from left to right are normalized $Vp$, gradients of $Vp$, and the Laplacians of $Vp$; rows from top to bottom correspond to ground truth, representations using a SIREN, and representations using a ReLU-based MLP.}
	\label{fig:siren-relu}
\end{figure}

In the subsurface parameterizations representation, the SIREN and ReLU-based MLP take approximately 5000 and 50000 training epochs, respectively, until their convergence occur. Since they are almost equally efficient in forward and backward propagations, the SIREN exhibits a convergence rate ten times faster than the ReLU-based MLP. The final representations using well trained networks are shown in Figure~\ref{fig:siren-relu}. In the first row of Figure~\ref{fig:siren-relu}, we compute the normalization, the gradients, and the Laplacian of $Vp$ using a grid-based Marmousi model. The second and third rows of Figure~\ref{fig:siren-relu} show implicit representations of Marmousi and their gradients and Laplacian computed using the SIREN and the ReLU-based MLP, respectively. By comparing the first column of Figure~\ref{fig:siren-relu}, we conclude that both networks have the ability to reconstruct primary structures of the subsurface. However, the SIREN outperforms the ReLU-based MLP in the representation of fine structures, as shown in the yellow box. A similar performance is observed in the second column of Figure~\ref{fig:siren-relu} that the SIREN is able to reconstruct the gradients of $Vp$ in finer detail, while the ReLU-based MLP can only provide a fraction of them. Moreover, in the representation of high-order derivatives shown in the third column of Figure~\ref{fig:siren-relu}, the SIREN provides the Laplacian of $Vp$ with a remarkable resolution, where the ReLU-based MLP completely fails. 

\begin{figure}[ht]
	\centering
	\includegraphics[width=\textwidth]{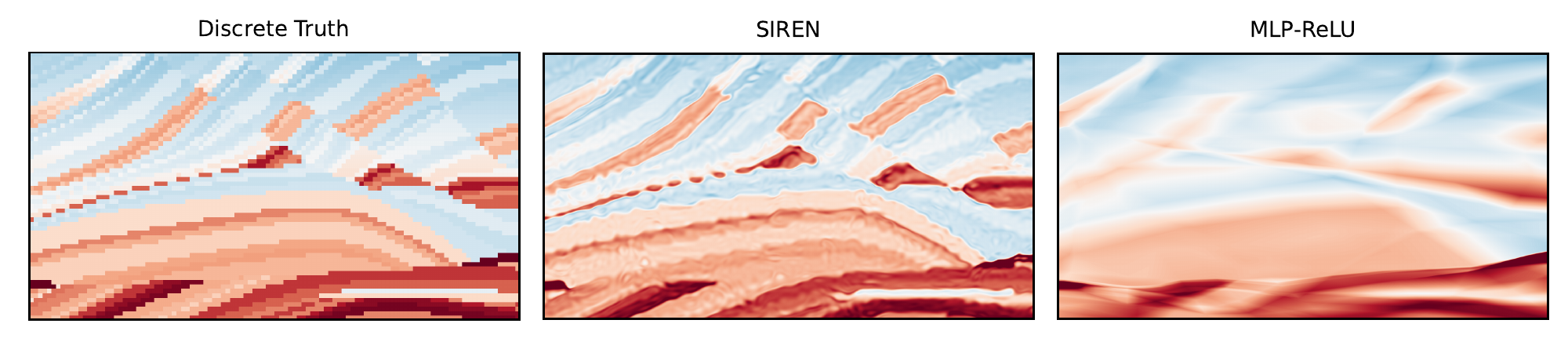}
	\caption{An enlarged view of the selected area indicated by the yellow box in Figure~\ref{fig:siren-relu}.}
	\label{fig:siren-trunc}
\end{figure}

In Figure~\ref{fig:siren-trunc}, we show the enlarged view of the selected region indicated by the yellow box in Figure~\ref{fig:siren-relu}, which further confirms the capability of the SIREN in fine structure representation. In addition, the serrated effects can be observed in the discrete grid-based parameterizations (the left panel of Figure~\ref{fig:siren-trunc}) due to the grid limited resolution. Figure~\ref{fig:siren-trunc} indicates that implicit representations using MLPs are able to exhibit a certain degree of continuity, although slight discontinuities can be observed in the SIREN representation (the middle panel of Figure~\ref{fig:siren-trunc}), which are likely caused by the overfitting and will be discussed later.

From the above experiments, we observe that the SIREN outperforms the ReLU-based MLP in representing subsurface parameterizations, including a faster convergence rate and the ability to outline fine structures and high-order derivatives. Therefore, the SIREN with the same architecture (i.e., containing four hidden layers with 128 neurons in each) is applied to the rest of the work in this paper. The input and output layers are determined by the number of input and output features, respectively. For instance, the input layer contains two neurons if the two-dimensional spatial coordinates are used as inputs; the output layer contains three neurons if we are reconstructing the parameterizations of density, compressional- and shear-wave velocities. It must be emphasized that the network architecture chosen in this paper may not be optimal and better performances may be achieved using other classic architectures, such as CNNs, vision Transformers, and graph neural networks (GNNs). However, in order not to deviate from the main line of IFWI, we leave it to readers to discover the best design for network architecture.

\subsection{Why IFWI works}

\begin{figure}[ht]
	\centering
	\includegraphics[scale=0.7]{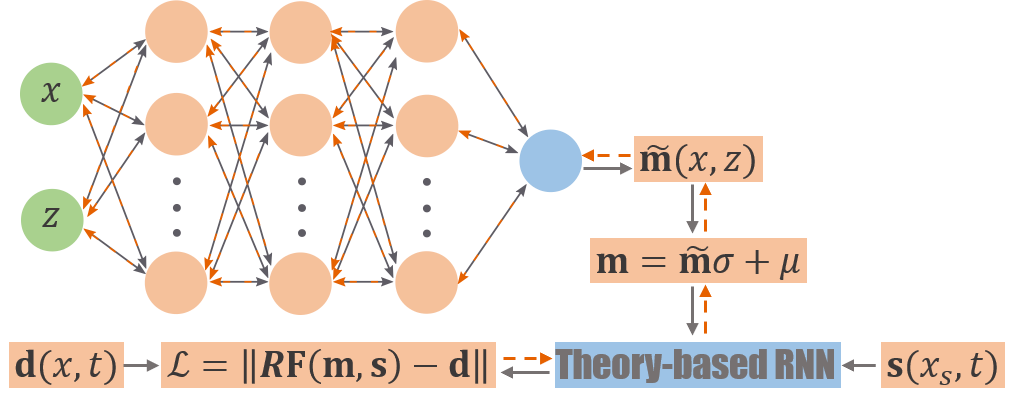}
	\caption{The workflow of implicit full waveform inversion with deep neural representation.}
	\label{fig:ifwi-workflow}
\end{figure}

With the success of DNR in subsurface representation, the workflow of IFWI is schematically shown in Figure~\ref{fig:ifwi-workflow}. First, we feed the spatial coordinates ($\mathbf{x}$ and $\mathbf{z}$) into the randomly initialized SIREN, and then perform the anti-normalization using a mean and a standard deviation. The mean and standard deviation can be calculated using the well-log information. Without any prior knowledge, one can use a global mean ($\mu=3.0$ km/s) and a global standard deviation ($\sigma=1.0$ km/s). We believe that the global mean and standard deviation are good enough for most subsurface models, as long as they cover the minimum and maximum values of the model to be represented, since they are only used to shift and scale the distribution of the output features. In the following contexts, we use the global mean and standard deviation for all Marmousi-related experiments. Second, the parameterizations constructed by the SIREN are employed to perform the forward modeling for simulated data generation. Third, we compute the discrepancy between simulated data and observation and backpropagate it to update weights of SIREN. Next, repeat all above steps until convergence.

Geophysical inversion can be delineated as finding a unique model that fits data acceptably. The fact that we can only observe the bandlimited data makes geophysical inversion a non-convex optimization problem and generally suffer from the non-uniqueness and local minima. Thus, good initial models or low-frequency components are often required in the procedure of geophysical inversion and FWI. However, we argue that IFWI works with a randomly initialized model without pretraining the network and the same scheme can be easily generalized in other kinds of geophysical inversion. In the following, we demonstrate this statement from both theoretical and experimental aspects.

Theoretically, the reason for this is that a critical point, where the gradient at this location is zero, is unlikely to be a local minimum in deep learning optimization. It is well known that a critical point $\mathbf{x} \in \mathbb{R}^n$ in the $n$ dimension to be a local minimum is that the Hessian matrix $\mathbf{H}(\mathbf{x})$ at this point must be positive semi-definite. Due to its symmetric characteristics, the Hessian at this critical point can be diagonalized with the following mathematical expression:
\begin{equation}
  \mathbf{H}(\mathbf{x}) =
  \begin{bmatrix}
    \lambda_{1} & & \\
    & \ddots & \\
    & & \lambda_{n}
  \end{bmatrix}
\end{equation}
where all elements of $\mathbf{H}(\mathbf{x})$ are non-negative, i.e., $\lambda_i \geqslant 0$ for $0 \leqslant i \leqslant n$.

Considering the high dimensional non-linearity of the Hessian, we assume that the signs of the elements in $\mathbf{H}(\mathbf{x})$ are independent and the probability of each of them being non-negative is $1/\kappa$ with $\kappa > 1$. Thus, the probability of a given critical point being a local minimum can be expressed as
\begin{equation} \label{eqn:prob}
P(\mathbf{x}_{min}) = P(\lambda_1 \geqslant 0, \lambda_2 \geqslant 0, \dots, \lambda_n \geqslant 0) = \prod_{i=1}^{n}
P(\lambda_i \geqslant 0) = \frac{1}{\kappa^n}
\end{equation}

Equation~\ref{eqn:prob} states that the probability of a critical point being a local minimum decreases exponentially with increasing dimensionality. The DNN usually consists of more than million of trainable parameters, which define an ultra-high dimensional space (i.e., $n > 10^6$ ). In other words, a critical point in the deep learning optimization is most likely a saddle point, instead of a local minimum, i.e., $\frac{1}{\kappa^n} \rightarrow 0$. Note that this is not a strict proof and more theoretical analysis about convergence of deep learning are investigated based on landscape conjecture \citep{choromanska2015loss,bhojanapalli2016global,ge2015escaping,ge2016matrix} or trajectory \citep{brutzkus2017globally,brutzkus2017sgd,li2017convergence,tian2017analytical,arora2018convergence}. However, it is fair to say that IFWI may greatly reduce the non-uniqueness of inversion results and the risk of falling into local minima during the optimization.

From an empirical point of view, let's use a primitive example to illustrate the nature of the neural network learning process. Suppose we want a neural network to learn a continuous signal, such as an oscillating sinusoidal function, a 1D seismic trace, and a 1D velocity profile. A SIREN is adopted and its weights are optimized by minimizing the misfits between mimic signals and the ground truth. We record the learning process of the neural network in tasks of representing the three signals mentioned above, in which the oscillated sinusoidal signal is specifically designed as a combination of a low-frequency sinusoidal signal and a high-frequency sinusoidal signal.

\begin{figure}[ht]
	\centering
	\includegraphics[width=\textwidth]{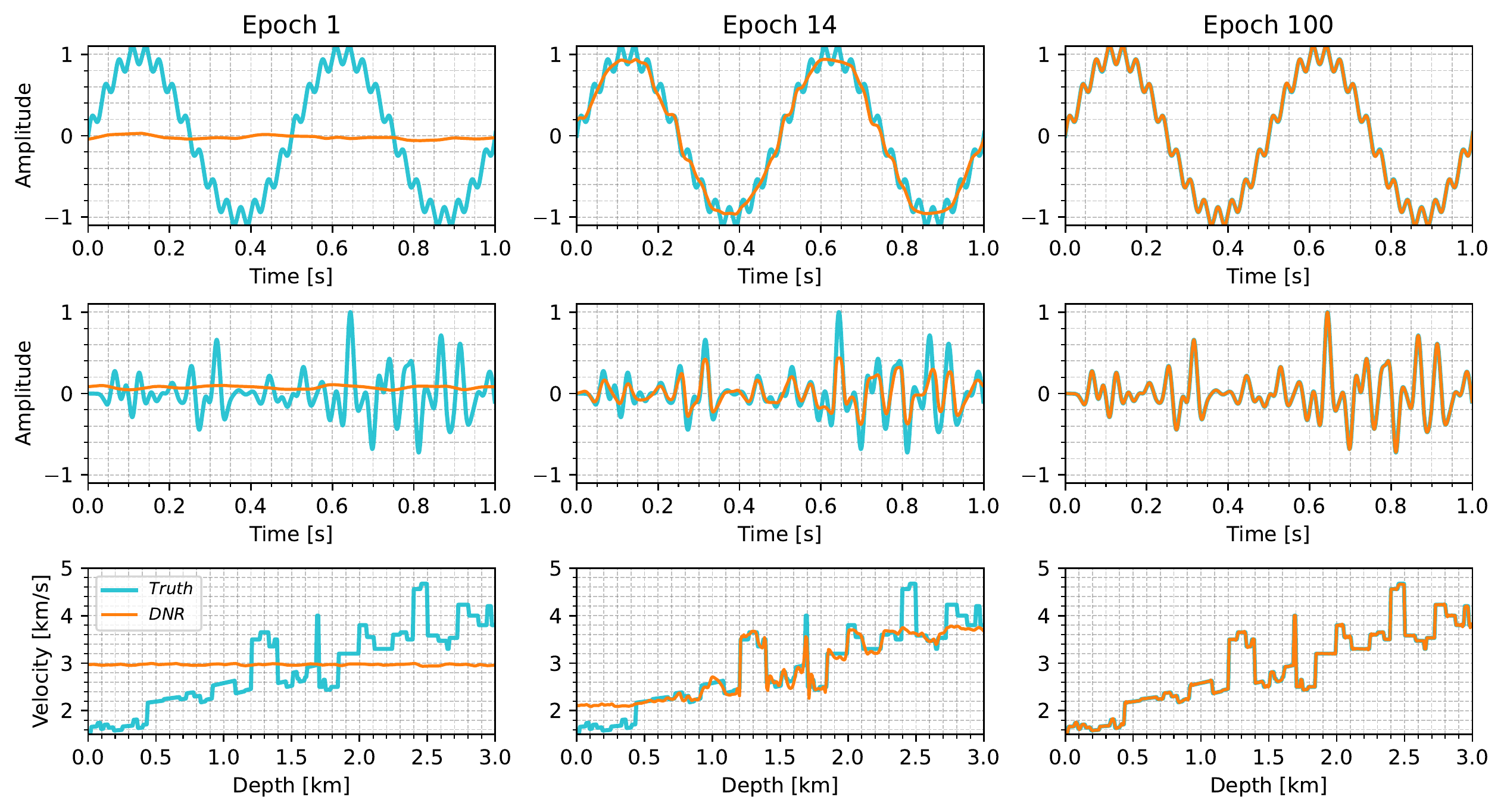}
	\caption{Frequency bias of MLPs with sinusoidal activation functions.}
	\label{fig:freqbias}
\end{figure}

Figure~\ref{fig:freqbias} shows the learning process of the network representing these three signals and the comparison between learned and ground-truth signals at three different training epochs. During the training of neural networks, we observe that low-frequency information are always learned first compared to high-frequency information. This phenomenon is known as spectral/frequency bias \citep{rahaman2019spectral,basri2020frequency}. For simplicity, we refer to it as frequency bias in the following contexts. It indicates that frequency bias may ensure IFWI firstly updates the low-wavenumber components, rather than the high-wavenumber components, of subsurface models. Thus, both theoretical and empirical results indicate that IFWI has the ability to converge using a random initialization.

\section{Numerical examples}
\begin{figure}[ht]
	\centering
	\includegraphics[scale=0.7]{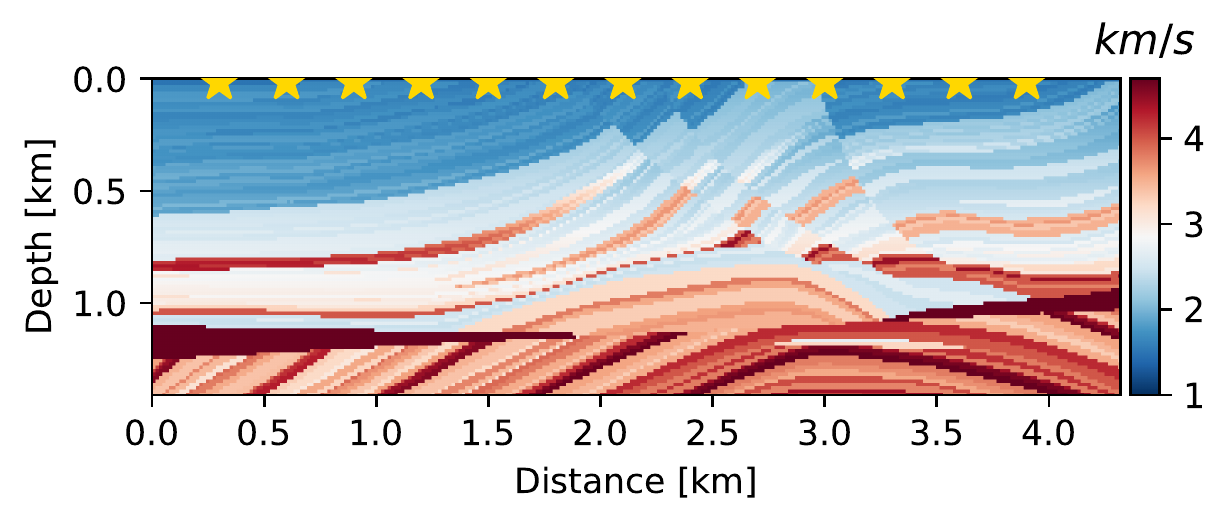}
	\caption{Compressional wave velocity of Marmousi model, and source locations are indicated by stars.}
	\label{fig:marmousi}
\end{figure}

\begin{figure}[ht!]
	\centering
	\includegraphics[scale=0.6]{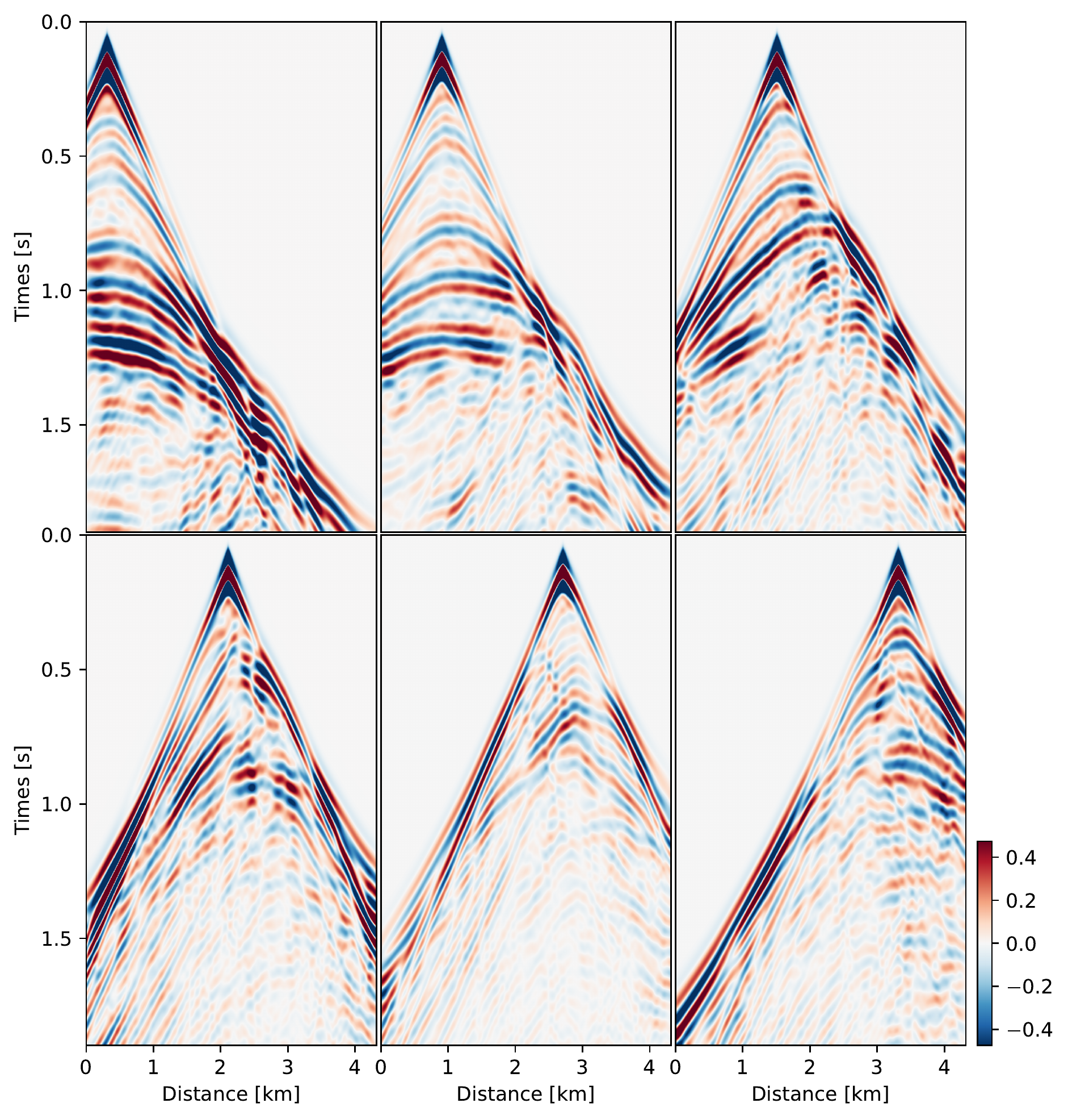}
	\caption{Six representative examples of 13 shot gathers.}
	\label{fig:shots}
\end{figure}

In this section, we further examine the effectiveness and potentials of IFWI using a 2D acoustic Marmousi model with a grid size $94\times288$ and 15m cells, shown in Figure~\ref{fig:marmousi}. 13 synthetic shot gathers are collected with a shot interval of 300m at a depth of 30m using a Ricker wavelet with 8Hz dominant frequency, where shot locations are indicated as the golden stars in Figure~\ref{fig:marmousi}. A free surface at the top and a perfectly matched layer (PML) absorbing condition at three other boundaries are applied during the synthetic seismic record collection. The time sample interval during wave propagation is 1.9ms, and 1000 time samples are recorded. Figure~\ref{fig:shots} shows six representative examples of 13 shot gathers collected using Marmousi model. To ensure the fairness of all comparisons, both FWI and IFWI are performed with a theory-designed RNN framework using their respective optimal learning rates.

\subsection{Random initialization}
\begin{figure}[ht!]
	\centering
	\includegraphics[scale=0.65]{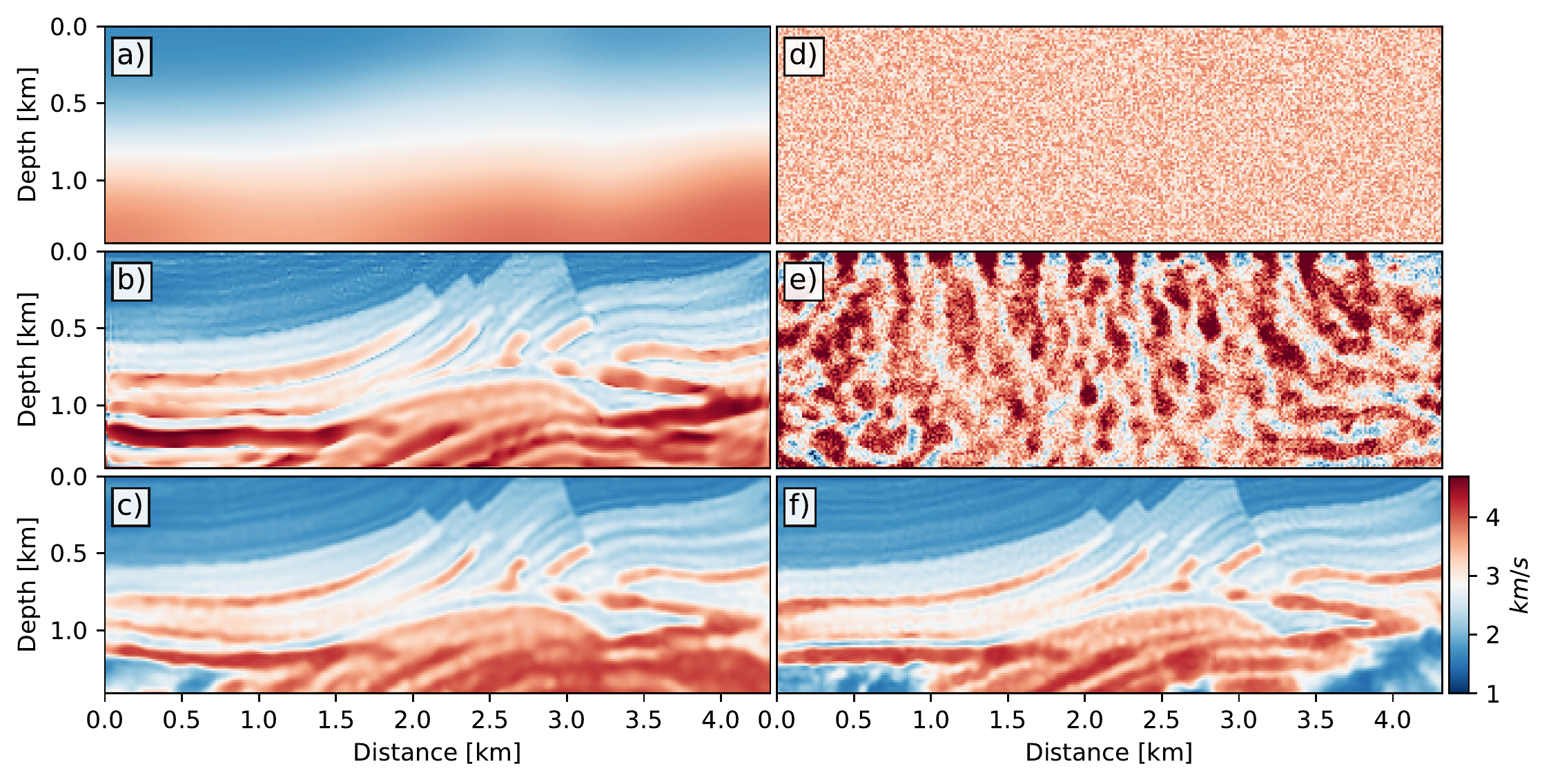}
	\caption{Comparisons of FWI and IFWI with a smooth initial model and a random initial model, respectively. a) the smooth initial model, b) FWI results with the smooth initial model shown in Figure~\ref{fig:comparisons_fwi_ifwi}a, c) IFWI results by pretraining the network with the smooth initial model shown in Figure~\ref{fig:comparisons_fwi_ifwi}a, d) the random initial model, e) FWI results with the random initial model shown in Figure~\ref{fig:comparisons_fwi_ifwi}d, f) IFWI results with the random initial model shown in Figure~\ref{fig:comparisons_fwi_ifwi}d.}
	\label{fig:comparisons_fwi_ifwi}
\end{figure}

For a fair comparison with FWI, we first perform both FWI and IFWI with a smooth initial model, which is obtained by applying a Gaussian filter on the ground truth model shown in Figure~\ref{fig:marmousi}. The computed smooth initial model is plotted in Figure~\ref{fig:comparisons_fwi_ifwi}a. For the implementation of FWI with a smooth initial model (refer to as FWI-Smooth), an optimal learning rate of 0.01 is applied based on the instructive analysis of \cite{sun2020theory}, and the final inversion result of FWI is plotted in Figure~\ref{fig:comparisons_fwi_ifwi}b. For IFWI implementation, the smooth initial model shown in Figure~\ref{fig:comparisons_fwi_ifwi}a is employed to pretrain the SIREN, which is then embedded into the IFWI procedure. With trial-and-error experiments, a same learning rate of 0.0001 is adopted for both pretraining and IFWI processes. The final inversion result of IFWI using a SIREN pretrained by a smooth initial model (refer to as IFWI-Pretrain) is plotted in Figure~\ref{fig:comparisons_fwi_ifwi}c. As indicated in the first column of Figure~\ref{fig:comparisons_fwi_ifwi}, given a good initial model, both IFWI and FWI have the ability to converge to the global minimum and produce satisfactory results.

In addition, another experiment is designed to examine the performance of IFWI with randomly initialized starting model. The initial model shown in Figure~\ref{fig:comparisons_fwi_ifwi}d is randomly generated using a normal distribution and then is anti-normalized with the global mean and standard deviation. Both FWI and IFWI with this random initialization (refer to as FWI-Random and IFWI-Random, respectively) are performed using their respective optimal learning rates (0.001 and 0.0001, respectively) and their final results are plotted in Figure~\ref{fig:comparisons_fwi_ifwi}e and \ref{fig:comparisons_fwi_ifwi}f, respectively. Figure~\ref{fig:comparisons_fwi_ifwi}e indicates that, without a good initial model, FWI can easily fall into local minima during the optimization process and cannot predict the adequate velocity model. Compared to FWI, Figure~\ref{fig:comparisons_fwi_ifwi}f shows that IFWI is able to reconstruct satisfactory results containing fine detailed structures even when starting from a random initial model, which further validates our inference about IFWI. By further comparing Figure~\ref{fig:comparisons_fwi_ifwi}b and \ref{fig:comparisons_fwi_ifwi}f, we learn that IFWI with a random initialization can produce comparable results to FWI with a good initial model.

\begin{figure}[ht]
	\centering
	\includegraphics[scale=0.7]{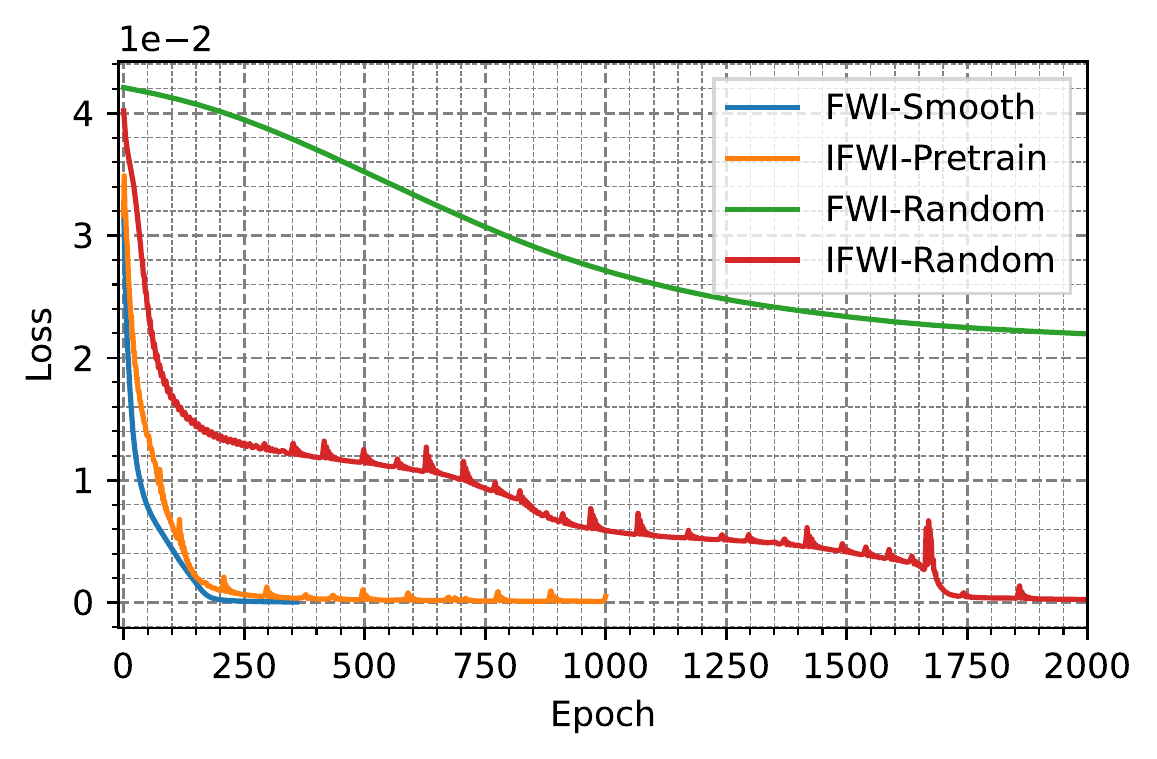}
	\caption{The variations of data discrepancies during four implementations, including FWI with a smooth initial model (FWI-Smooth), IFWI with a SIREN pretrained by a smooth initial model (IFWI-pretrain), FWI with a random initial model (FWI-Random), and IFWI with a random initial model (IFWI-Random), and their corresponding final results are plotted in Figure~\ref{fig:comparisons_fwi_ifwi}b, \ref{fig:comparisons_fwi_ifwi}c, \ref{fig:comparisons_fwi_ifwi}e, and \ref{fig:comparisons_fwi_ifwi}f, respectively.}
	\label{fig:training_losses}
\end{figure}

Figure~\ref{fig:training_losses} shows the relative computational costs of four implementations mentioned previously, including FWI-Smooth, FWI-Random, IFWI-Pretrain, and IFWI-Random. We observe that, when a good initial model is given, both FWI-Smooth and IFWI-Pretrain are able to converge quickly and with comparable efficiency, as delineated in blue and orange lines in Figure~\ref{fig:training_losses}. When lacking a good initial model, it is difficult for FWI to escape from the trap of local minima, however, IFWI can gradually converge to the global minimum. Understandably, IFWI-Random takes longer to converge, as it starts from a random initial model. Note that better converge rate may be acquired with an alternative neural network architecture.

\begin{figure}[ht]
	\centering
	\includegraphics[scale=0.65]{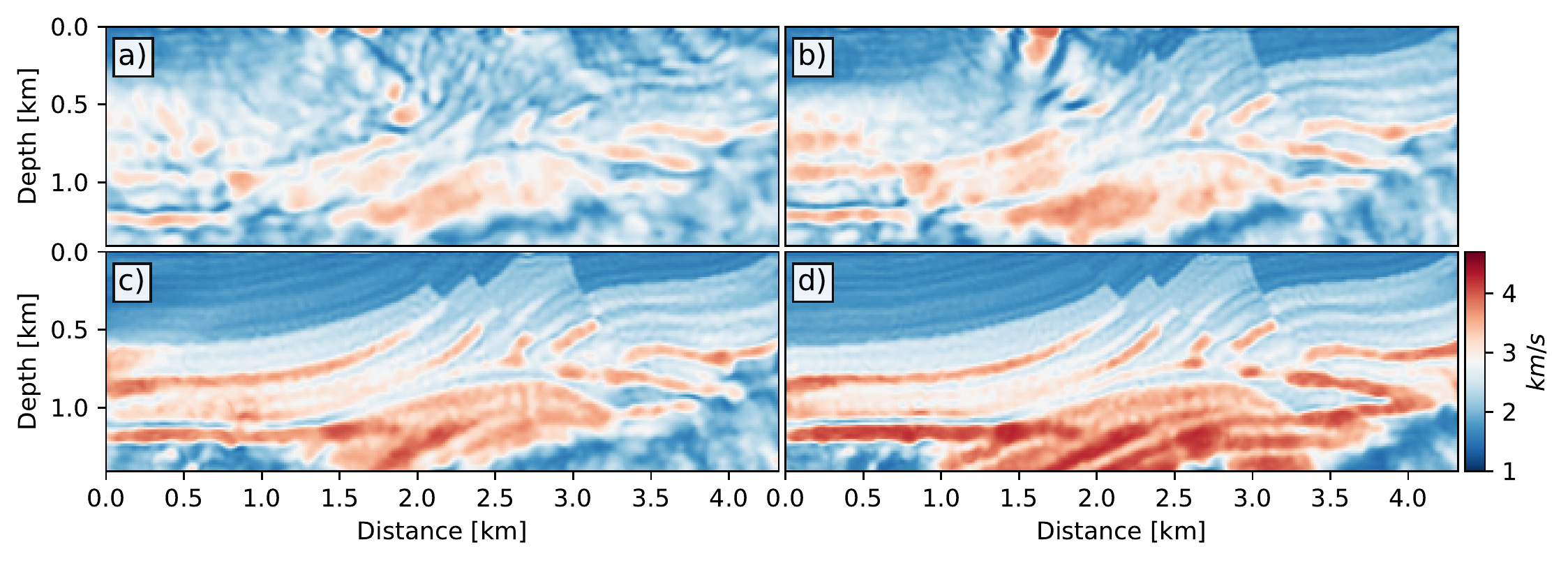}
	\caption{The iterative inversion results of IFWI. a) the inversion at 500th epoch, b) the inversion at 1000th epoch, c) the inversion at 1500th epoch, and d) the inversion at 2000th epoch.}
	\label{fig:ifwi_train}
\end{figure}

To analyze the inversion process of IFWI more intuitively, we plot inversion results at 500th, 1000th, 1500th, 2000th training epochs in Figure~\ref{fig:ifwi_train}.  The low-wavenumber information of the Marmousi model are evidently recovered after 500 epochs. With 1000 epochs, IFWI is able to recover primary structures with robust layer information in shallow zones. After 1500 epochs, the stratigraphy of shallow zones are refined with detailed information, and some of deep structures also appear. IFWI is able to further improve the precision of subsurface parameters and layers in deep zones after 2000 epochs. The inversion process of IFWI on the 2D Marmousi model is also consistent with our conclusions and experiments on frequency bias in the previous section. This set of experiments shows that it is not only feasible for IFWI to represent the subsurface parameters by a continuous implicit neural network, but it also allows for a stochastic initial model due to the increased degrees of freedom.

\subsection{Robustness}
\begin{figure}[ht!]
	\centering
	\includegraphics[scale=0.6]{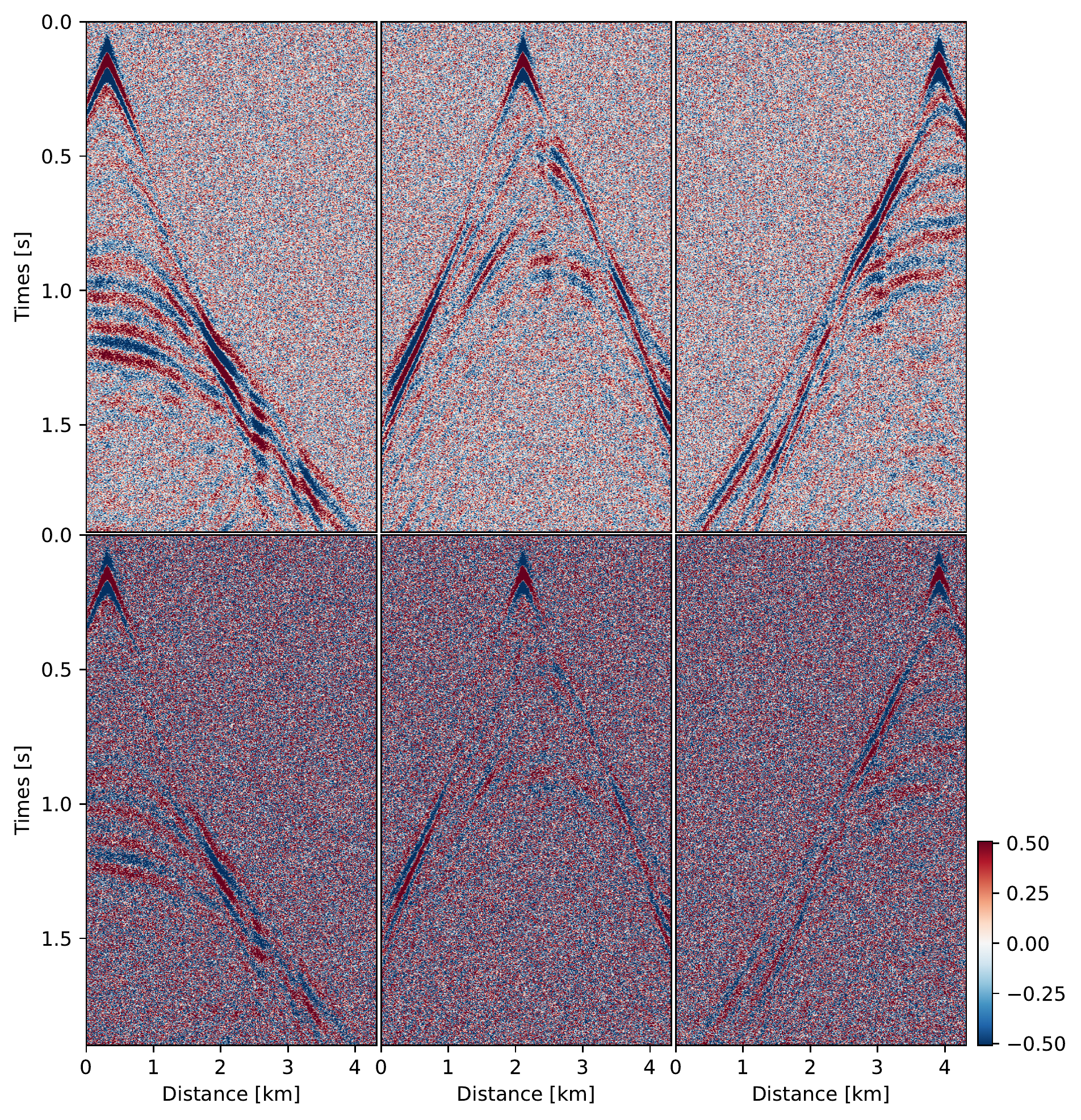}
	\caption{Representative examples of noisy shot gathers. Top row: 3 representative shots with random noise of standard deviation $\sigma=2\sigma_0$, bottom row: 3 representative shots with random noise of standard deviation $\sigma=4\sigma_0$.}
	\label{fig:shots_noisy}
\end{figure}

Next, we exemplify the robustness of IFWI algorithm by adding different levels of random noise into the observed data. Gaussian white noises are generated using standard deviations $\sigma = 2 \sigma_0$ and $\sigma = 4 \sigma_0$, respectively, where $\sigma_0$ represents for the standard deviation of the noise-free observed data. In Figure~\ref{fig:shots_noisy}, three representative shot gathers with random noise of standard deviation $\sigma=2\sigma_0$ (refer to as noisy-data $\sigma=2\sigma_0$) are plotted in the top row, and data with random noise of standard deviation $\sigma=4\sigma_0$ (refer to as noisy-data $\sigma=4\sigma_0$) are plotted in the bottom row. For comparison purposes, both FWI and IFWI are performed using noisy observations as the ground truth signals in the calculation of data discrepancy, where the smooth initial model (Figure~\ref{fig:comparisons_fwi_ifwi}a) is employed during the implementation of FWI and IFWI utilizes the random initialization (Figure~\ref{fig:comparisons_fwi_ifwi}d) as the starting model. 

\begin{figure}[ht!]
	\centering
	\includegraphics[scale=0.7]{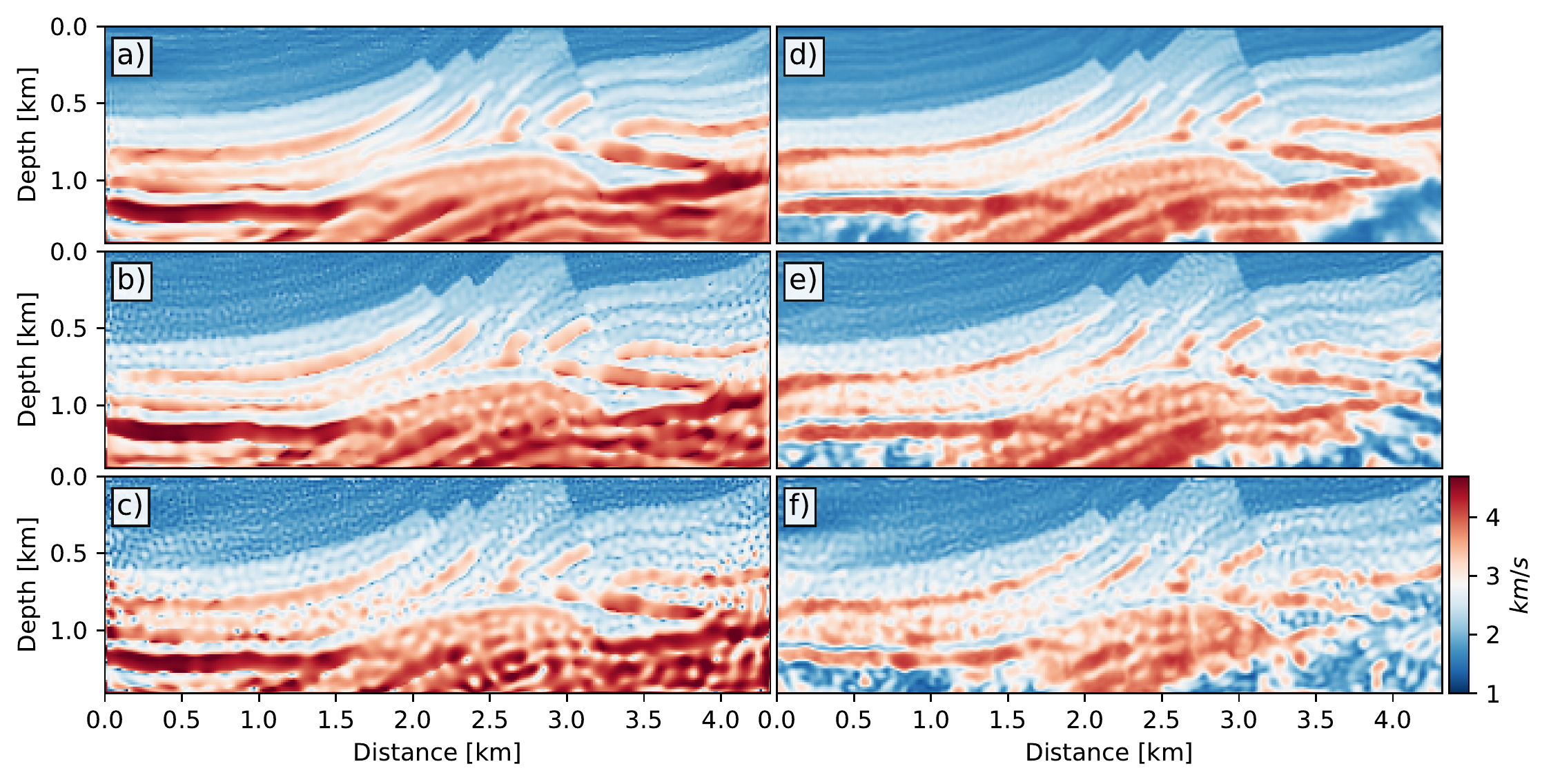}
	\caption{Comparisons of inversion results for FWI and IFWI using noise-free and noisy observations. a) FWI with noise-free data, b) FWI with noisy-data $\sigma=2\sigma_0$, c) FWI with noisy-data $\sigma=4\sigma_0$ , d) IFWI with noise-free data, e) IFWI with noisy-data $\sigma=2\sigma_0$, f) IFWI with noisy-data $\sigma=4\sigma_0$.}
	\label{fig:comparison_noisy}
\end{figure}

Comparisons of final inversion results are shown in Figure~\ref{fig:comparison_noisy}, where FWI and IFWI are plotted from left to right by column, and the noise levels are differentiated in a row perspective. Specifically, inversion results using noise-free data, noisy-data $\sigma=2\sigma_0$, and noisy-data $\sigma=4\sigma_0$ are plotted in the first, second, and third rows in Figure~\ref{fig:comparison_noisy}. Figure~\ref{fig:comparison_noisy}b and ~\ref{fig:comparison_noisy}c indicate that, given a proper initial model, FWI exhibits great tolerance to noise, although some small perturbations are observed in the final results. Compared to FWI, IFWI shows considerable robustness capability even with a random initial model, which are confirmed in Figure~\ref{fig:comparison_noisy}e and \ref{fig:comparison_noisy}f. In this robustness experiments, both FWI and IFWI are implemented in a regularization-free manner, therefore, better results may be achieved if some form of regularization is added.

\begin{figure}[ht!]
	\centering
	\includegraphics[scale=0.7]{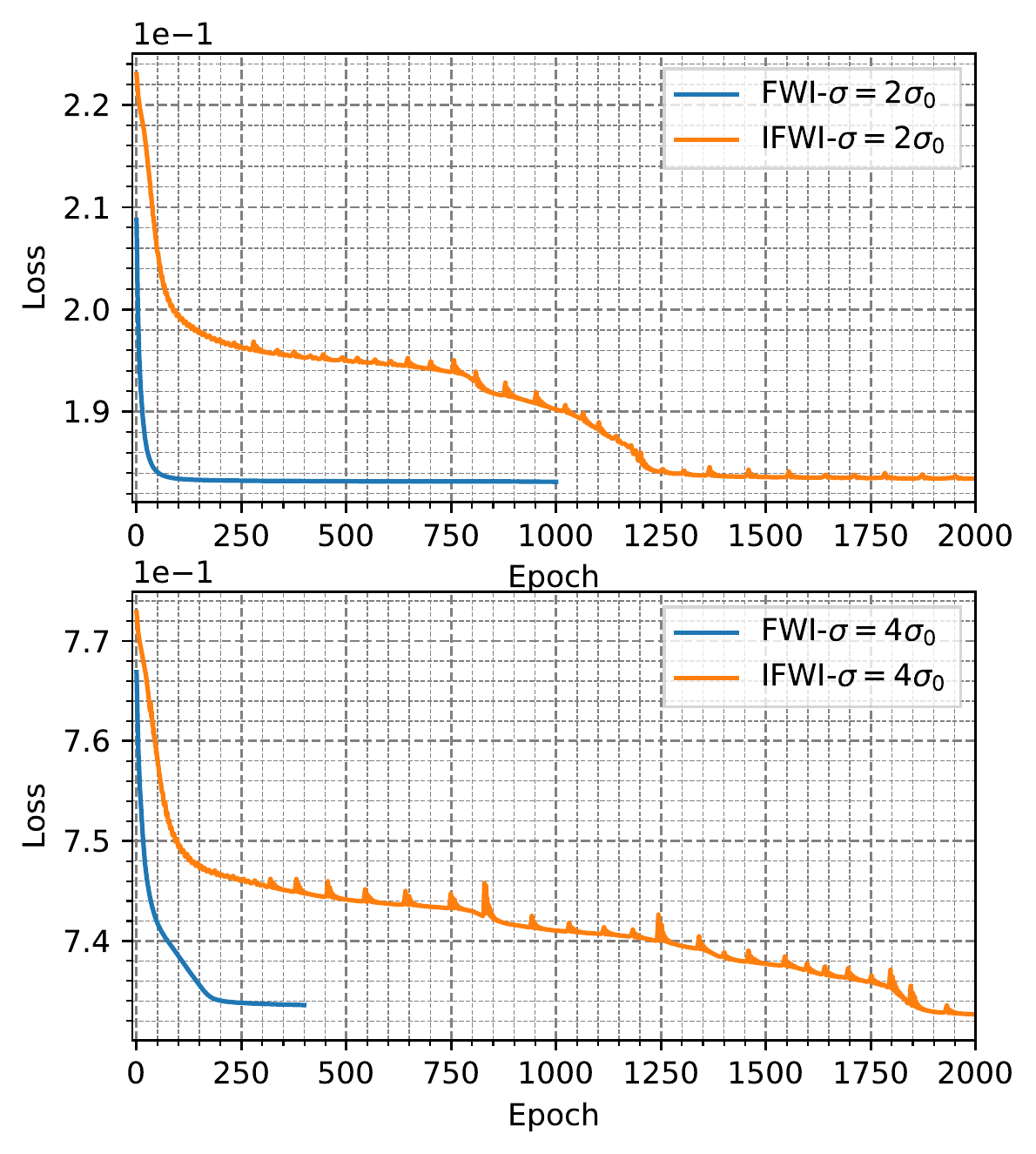}
	\caption{The variations of data discrepancies during FWI and IFWI implementations with noisy data. Top: losses using noisy data with $\sigma = 2\sigma_0$, bottom: losses using noisy data with $\sigma = 4\sigma_0$.}
	\label{fig:training_losses_noisy}
\end{figure}

In order to analyze the convergence process from a macroscopic point of view, we plot the monitored objective losses with noisy observations for FWI and IFWI in Figure~\ref{fig:training_losses_noisy}. The top panel of Figure~\ref{fig:training_losses_noisy} shows data discrepancies with noisy-data $\sigma=2\sigma_0$ versus training epoch, in which we observe slightly faster convergence rates for both FWI and IFWI than their corresponding noise-free implementations, i.e., FWI-Smooth and IFWI-Random delineated as blue and red lines in Figure~\ref{fig:training_losses}, respectively. The objective losses of data discrepancies with noisy-data $\sigma=4\sigma_0$ versus training epoch are plotted in the bottom panel of Figure~\ref{fig:training_losses_noisy}, and their convergence rates are roughly equivalent to their noise-free cases. Due to the robustness of IFWI, a trade-off between the convergence rate and the accuracy of the inversion result may be achieved by adding a small amount of random noise.
 
\subsection{Uncertainty}
One of the most challenge tasks in FWI is the uncertainty assessment of the inversion results due to the complexity and prohibitive computational cost of calculating the Hessian or posterior covariance matrix for a large amount of subsurface parameters. Though the supervised deep learning aims to establish an end-to-end relationship from the input to output, which is generally deterministic, there are still several approaches to approximate the Bayesian inference. For instance, deep ensemble learning can produce collective predictions that approach Bayesian predictive distribution by retaining the same neural network multiple times using the respective data sets \citep{lakshminarayanan2017simple}. Deep ensembles can be applied to IFWI framework if abundant observations are available. However, independently carrying out IFWI multiple times is a very time-consuming task. Another alternative is to use a Bayesian neural network \citep[i.e., BNN,][]{kononenko1989bayesian, jospin2022hands} to represent subsurface parameters, in which weights and biases are randomly generated from a probability distribution trained by IFWI. Thus, instead of a deterministic implicit representation of subsurface parameters, the BNN-based IFWI can produce the posterior distribution of subsurface model with given datasets. By evaluating the trained BNN multiple times, we can obtain collective inversion results of subsurface models, which allows us to estimate uncertainty in predictions. In order to concentrate the concept of IFWI and to keep this paper concise, the theory of BNN and BNN-based IFWI are not further discussed in this paper. Readers interested in BNN can refer to the work of \cite{jospin2022hands}. 

Besides the deep ensembles and BNN, the third, and simplest, way to gain Bayesian posterior distribution of predictions is to include dropout neurons in both training and validating procedures. \cite{gal2016dropout} demonstrate that training the neural network with dropouts activated is theoretically equivalent to training a large ensemble of networks, which approach to a Bayesian statistical model. Though dropout operation is initially introduced to prevent networks from overfitting as a regularization method during training process \citep{srivastava2014dropout}, and is usually deactivated for the deterministic prediction in validation, the predictive realizations can also be obtained by validating the trained neural network multiple times while activating dropouts \citep{gal2016dropout}. \cite{sun2021physics} firstly introduced the dropout method to evaluate the predictive uncertainty of seismic inversion. Taking advantage of its simplicity, we give the example of IFWI in predictive uncertainty analysis by easily adding the dropout neurons to the representation network.

\begin{figure}[ht!]
	\centering
	\includegraphics[scale=0.7]{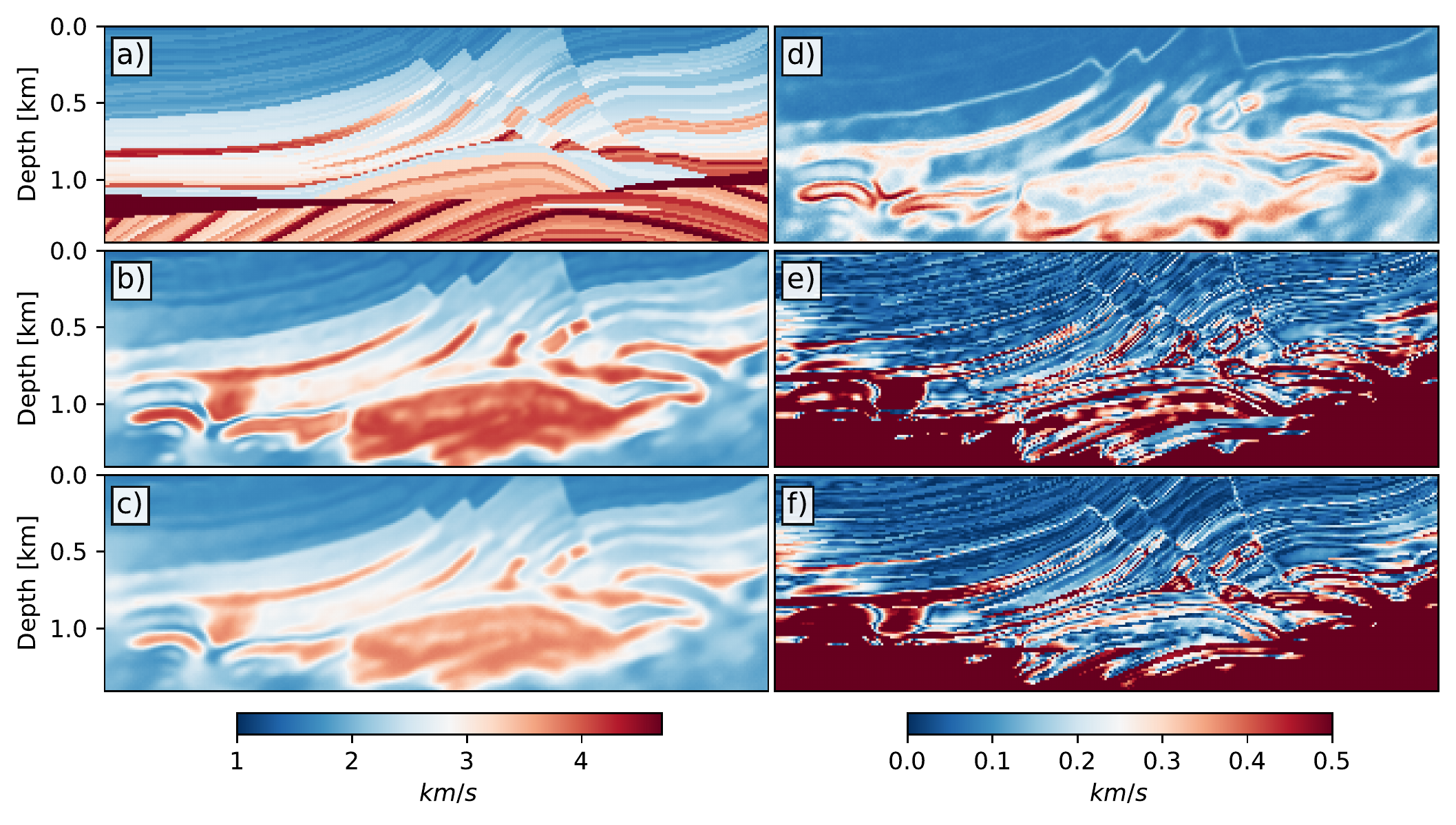}
	\caption{The uncertainty measurements of IFWI with dropout ratio $p=0.2$. a) the ground-truth Marmousi model, b) the inversion result without activated dropout in validation, c) the average map of 1000 realizations, d) the standard deviation of 1000 realizations, e) the absolute error map between Figures~\ref{fig:ifwi_uncertainty}a) and ~\ref{fig:ifwi_uncertainty}b), f) the absolute error map between Figures~\ref{fig:ifwi_uncertainty} a) and ~\ref{fig:ifwi_uncertainty}c).}
	\label{fig:ifwi_uncertainty}
\end{figure}

In every training epoch, we randomly selected $80\%$ of neurons of hidden layers using a Bernouli distribution to form a new neural network, where the remaining $20\%$ (i.e., dropout ratio $p=0.2$) of neurons are reset to zero. In validation, we first carry out the prediction using the trained network without activating dropout neurons, which are commonly considered the optimal generative performance of the network. The final prediction and its absolute misfits are plotted in Figures~\ref{fig:ifwi_uncertainty}b and \ref{fig:ifwi_uncertainty}e, respectively. Compared to IFWI without dropouts (refer to as non-dropout IFWI, shown in Figure~\ref{fig:ifwi_train}d), Figure~\ref{fig:ifwi_uncertainty}b indicates that training IFWI with dropouts may produce smoother inversion results, which is equivalent to adding certain forms of regularization into the non-dropout IFWI during optimization. However, we observe the presence of missing inversions in some areas, especially in the deep corners, which is likely caused by the difficulty of training networks with dropouts. This can usually be compensated for by either increasing the length of training time or increasing the complexity of the network.

To evaluate the predictive uncertainty of inversion results, we run 1000 realizations using the trained network by IFWI with dropouts activated. The mean and standard deviation of 1000 realizations are calculated and plotted in Figure~\ref{fig:ifwi_uncertainty}c and \ref{fig:ifwi_uncertainty}d, respectively. The averaged inversion result shown in Figure~\ref{fig:ifwi_uncertainty}c exhibits an analogous depiction of subsurface structures, but with moderate accuracy of velocity values, where the absolute misfits between the averaged result and the ground-truth are delineated in Figure~\ref{fig:ifwi_uncertainty}f. Note that the measured uncertainty map in Figure~\ref{fig:ifwi_uncertainty}d is not equivalent to the absolute error map shown in Figure~\ref{fig:ifwi_uncertainty}f, but shows a similar pattern. This is reasonable because the standard deviation represents for statistical biases of multiple realizations, where the absolute error map is calculated from a single realization.

\subsection{Generalization}
\begin{figure}[ht!]
	\centering
	\includegraphics[scale=0.7]{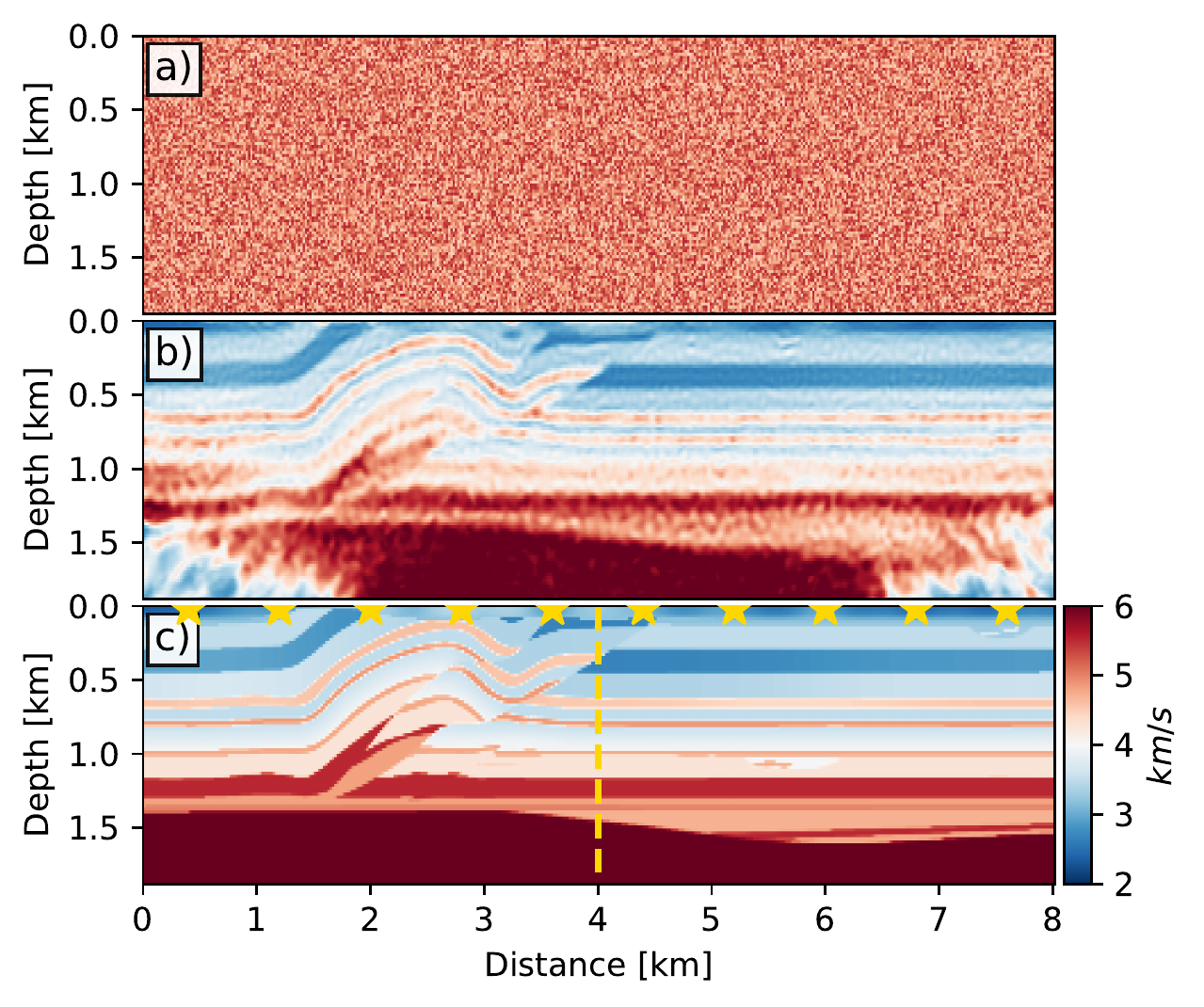}
	\caption{IFWI prediction of the 2D Overthrust model. a) the random initial model, b) the final inversion result, c) the ground-truth velocity with source locations indicated as golden stars and the well location indicated as the golden dashed line.}
	\label{fig:ifwi_overthrust}
\end{figure}

In order to examine the ability of IFWI to be applied in more general circumstances, we use a 2D slice extracted from the 3D Overthrust volume. The 2D Overthrust model is shown in Figure~\ref{fig:ifwi_overthrust}c with the grid cell of 20m. 10 shot gathers are acquired with 800m intervals at a depth of 20m, and their relative locations are indicated by the golden stars in Figure~\ref{fig:ifwi_overthrust}c. A Ricker wavelet with 8Hz dominant frequency is applied as the source and all shot gathers are sampled in 2ms with a length of 1500 samples. The free-surface at the top and a PML absorbing boundary condition at three other boundaries are applied during synthetic shot simulation. In this experiment, instead of using the global mean and standard deviation, we obtain a mean ($\mu=4.412km$) and a standard deviation ($\sigma=1.116km$) using a known well log indicated as the golden dashed line in Figure~\ref{fig:ifwi_overthrust}c. The implementation of IFWI on the 2D Overthrust model is also performed with a random initial model, shown in Figure~\ref{fig:ifwi_overthrust}a, using the same hyperparameters for optimization. Figure~\ref{fig:ifwi_overthrust}b shows the final inversion result of the 2D Overthrust model using IFWI. Precise fault structures and thin layers of the 2D Overthrust model are correctly inverted, which further confirm a strong generalization ability of IFWI algorithm.

\section{Discussions of potentials and challenges}
Our analysis confirms that IFWI has the advantages of being insensitive to the initial model, having a certain degree of robustness, and allowing intuitive uncertainty analysis of the inversion results, all of which can be applied to more general cases. Despite of these revealed benefits, more potentials of IFWI may be further explored. In this section, we discuss farther potentials of IFWI and the challenges that may encountered during the exploration. 

As our experiments have shown, IFWI allows to generate a coordinate-based and continuous representation of the subsurface model, which offers a golden opportunity for multiscale joint inversion, mesh-free and target-oriented inversion, and scalable subsurface model inversion. Specifically speaking, with a coordinate-based neural network, one can build a single DNR of subsurface models at different scales. The multiscale joint inversion can be performed by optimizing this single DNR with complementary constraints from different kinds of geophysical measurements, such as seismic, well-log, gravity, electromagnetic, ground penetrating radar, and remote sensing. In addition, one can implement the inversion in a fully mesh-free and target-oriented manner by means of suitable forward modeling algorithms, such as the PINN solver. Moreover, it is possible to perform IFWI on a coarse grid and then reconstruct the subsurface model in a fine grid, or perform IFWI on several discrete models and then generate a continuous subsurface model. However, there are challenges in undertaking such implementations. For example, we observe that ReLU-based MLPs can easily guarantee the continuity of the represented implicit functions, but have difficulties in learning their high-frequency components. MLPs with periodic activation functions \citep{sitzmann2020implicit} or Fourier features \citep{tancik2020fourier} are able to learn high-frequency information quickly, but they are also prone to overfitting, which can destroy the continuity of the represented spaces. Thus, a better network architecture is needed to balance its representation capacities in terms of continuity and high-frequency.

\section{Conclusions}
Geophysical inversions are great tools for reconstructing subsurface structures and estimating petrophysical properties, of which full waveform inversion (FWI) stands for the state-of-the-art technique. However, FWI usually suffers from local minima due to its strong non-linearity and is computationally demanding for uncertainty analysis. To address these issues, we propose the implicit full waveform inversion (IFWI) algorithm that produces a continuous and differentiable functional representation of subsurface parameters, instead of a grid-based solution. In contrast to the discrete parameterization, in which memory and precision are highly dependent on the grid solution, an implicit and continuous representation using deep neural networks (DNNs) can be much more efficient while preserving fine details. Both our theoretical and empirical analyses illustrate that IFWI is capable of automatic inversion from low- to high-frequencies through the frequency bias of deep learning, which may significantly reduce the reliance on the initial model. This is further confirmed by numerical examples using the 2D Marmousi model. Our experimentation indicates that, given a random initial model, IFWI can gradually converge to the global minimum and produce an impressive representation of subsurface parameters with fine structures, while FWI falls into a local minimum. In addition, uncertainty analysis can be easily performed during IFWI optimization by approximating Bayesian inference with various deep learning approaches, such as dropout neurons, Bayesian neural networks, and deep ensembles. Moreover, the robustness and generalization ability of IFWI are also exemplified by adding different levels of noise and using various geological models, respectively. Finally, we discuss further potentials and challenges of IFWI in multiscale and joint geophysical inversion.


\section*{Acknowledgment}
This research was supported by the Research Start-up Funding of Ocean University of China.
\bibliography{mybibfile}
	
\end{document}